\documentclass[a4paper]{jpconf}
\usepackage{graphicx}
\usepackage{amsmath, amssymb, amsfonts, bbm}
\usepackage{amssymb}

\newcommand{\sn}{\tilde{\nu}_{LSP}}
\begin{document}
\texttt{preprint: IFIC/12-63}\\

\title{Right Handed Sneutrino Dark Matter in Inverse and Linear seesaw scenarios}

\author{Valentina De Romeri}

\address{AHEP Group, Instituto de F\'\i sica Corpuscular -- 
C.S.I.C./Universitat de Val\`encia Edificio de Institutos de Paterna, 
Apartado 22085, E--46071 Val\`encia, Spain}

\ead{deromeri@ific.uv.es}
\begin{abstract}
We consider supersymmetric models in which the right$-$handed
sneutrino is a viable WIMP dark matter candidate. These are
either simple extensions of the Minimal Supersymmetric Standard Model or models with the addition of an extra U(1) group. All of them can explain small neutrino masses, through either the Inverse
or the Linear Seesaw mechanism. We investigate the properties
of the dark matter candidate naturally arising in these scenarios.
We check for phenomenological bounds, such as correct relic
abundance, consistency with direct detection cross section limits
and laboratory constraints. Especially, we comment on limitations
of the model space due to lepton flavour violating charged lepton decays.
\end{abstract}

\section{Introduction}
We analize two supersymmetric particle physics models, which contain a Right-Handed (RH) sneutrino $\sn$ as lightest supersymmetric particle (LSP) and thus as a good WIMP (weakly interactive massive particle) cold Dark Matter (CDM) candidate.\\
The two  models we consider are:\\  
\textbf{Model I} - A simple extension of the MSSM, which accounts for the {\it inverse seesaw} model \cite{Mohapatra:1986bd}, where  the particle content of the SM is extended by three pairs of singlets, called $\nu^c$ and $S$, which form ``heavy'' pseudo-Dirac pairs, or for an alternative seesaw scheme that can also be realized at low-scale: the {\it linear seesaw}, which has been suggested as arising from a particular $SO(10)$ unified model~\cite{Malinsky:2005bi,Akhmedov:1995ip,Akhmedov:1995vm}.\\
The total superpotential contains then the additional terms $M_R\,\hat{\nu}\,\hat{S}$,~ (Lepton Number - LN conserving) and $\frac{1}{2} \mu_S \,\hat{S}\,\hat{S}$,~ (LN violating):

\begin{align} 
W = & \,  Y_u\,\hat{u}^c\,\hat{q}\,\hat{H}_u\,- Y_d \,\hat{d}^c\,\hat{q}\,\hat{H}_d\,- Y_e \,\hat{e}^c\,\hat{l}\,\hat{H}_d\,+\mu\,\hat{H}_u\,\hat{H}_d\,+Y_\nu\,\hat{\nu}\,\hat{l}\,\hat{H}_u\,+M_R\,\hat{\nu}^c\,\hat{S}\,+\frac{1}{2} \mu_S \,\hat{S}\,\hat{S}\,
\label{eq:ModIsupp}
\end{align}

where $Y_e$, $Y_d$ and $Y_u$ are the usual MSSM Yukawa couplings for the charged leptons and the quarks. In addition there is the neutrino Yukawa coupling $Y_\nu$.\\
In the linear seesaw scheme, the superpotential is 

\begin{align} 
W = & \,  Y_u\,\hat{u}^c\,\hat{q}\,\hat{H}_u\,- Y_d \,\hat{d}^c\,\hat{q}\,\hat{H}_d\,- Y_e \,\hat{e}^c\,\hat{l}\,\hat{H}_d\,+\mu\,\hat{H}_u\,\hat{H}_d\,+Y_\nu\,\hat{\nu}\,\hat{l}\,\hat{H}_u\,+M_R\,\hat{\nu}^c\,\hat{S}\,+ Y_{SL}\,\hat{S}\,\hat{l}\,\hat{H}_u\,
\end{align} 

and Lepton Number is violated by $Y_{SL}$.

\textbf{Model II} -  A left-right symmetric extension of the SM based on a variant of the basic SUSY $SO(10)$ model advocated in \cite{Malinsky:2005bi} in which an extended intermediate
$U(1)_{R}\times U(1)_{B-L}$ stage follows a short $SU(2)_{R}\times U(1)_{B-L}$ phase, which can emerge, e.g.,  in
a class of  $SO(10)$ GUTs broken along the ``minimal'' left-right symmetric chain 
\begin{equation}
SO(10) \to  SU(3)_c\times SU(2)_L\times SU(2)_R \times U(1)_{B-L}
\to
   SU(3)_c\times SU(2)_L\times U(1)_R \times U(1)_{B-L},\nonumber
\end{equation}
in which the low energy scale seesaw schemes are actually naturally realized.\\

The  $R$-parity conserving superpotential is 
\begin{align}
W = & \,  Y_u\,\hat{u}^c\,\hat{q}\,\hat{H}_u\,- Y_d
\,\hat{d}^c\,\hat{q}\,\hat{H}_d\,+Y_\nu\,\hat{\nu}^c\,\hat{l}\,\hat{H}_u\,-
Y_e \,\hat{e}^c\,\hat{l}\,\hat{H}_d\,+Y_s\,\hat{\nu}^c\,\hat{\chi}_R\,\hat{S}\,+\mu\,\hat{H}_u\,\hat{H}_d\,-
\mu_R \,\hat{\bar{\chi}}_R\,\hat{\chi}_R\,+\frac{1}{2} \mu_S \,\hat{S}\,\hat{S}\,
\end{align}
with the same notation as before. The main introduction here is the term featuring the Yukawa coupling $Y_s$, which mixes the $\nu^c$ fields with the $S$ fields giving rise to an inverse seesaw mechanism for neutrino masses.
\section{Phenomenological constraints}
\subsection{Neutrino masses}
We fit neutrino masses according to neutrino oscillation data, i.e. with the atmospheric (atm) and solar ($\odot$) scales as in the case of normal hierarchy and the lepton mixing angles according to the most recent measurements of reactor antineutrino disappearance - Double CHOOZ, Daya Bay and RENO \cite{Abe:2011fz,An:2012eh,Ahn:2012nd,Tortola:2012te}
\subsection{Lepton flavor violation}
Putting all flavor in $Y_\nu$ is actually the usual option, thus keeping $Y_L$ , $M_R$ and $\mu_S$ diagonal. However, this framework gives strong constraints on the sneutrino LSP coming from LFV processes notably $\mu \longrightarrow e \gamma$ and three leptons decays like $\mu \longrightarrow e e e$.\\
Experiments set the bounds BR($\mu \longrightarrow e \gamma$) $\leq 2.4 \times 10^{-12}$ ($90 \%$ C.L.) \cite{Nishimura:2012zz,Adam:2011ch} and BR($\mu \longrightarrow e e e$) $\leq 1.0 \times 10^{-12}$ ($90 \%$ C.L.) \cite{Nakamura:2010zzi}. Indeed, it has been shown  \cite{Hirsch:2012ax} that in models with an extended particle content, $l_i \rightarrow 3 l_j$ can be more constraining than $l_i \rightarrow l_j \gamma$.\\
%Lepton flavor violation may be induced by the off-diagonal terms of the $M_R$ terms, although this setting leads to the same strong constraints as the previous case.\\
To really relax the LFV constraints, the best option is to put off diagonal terms in $\mu_S$ (in the case of the inverse seesaw) or in in $Y_{SL}$ (linear seesaw) and to keep $M_R$ and $Y_\nu$ diagonal
\subsection{Invisible Z width constraints}
Another experimental bound on MSSM neutrinos lighter than $m_Z/2$, comes from their contribution to the invisible $Z$--width, measured in searches at accelerators. Since in our models we have introduced also RH neutrinos $\nu^c$, the mixing (sin$\theta$) between the states has to be taken into account. The invisible $Z$--width including the mixing reads:
 \begin{equation}
\Delta\Gamma_{Z} = \sin^4\theta\,\frac{\Gamma_\nu}{2} \left[ 1- \left(\frac{2 m_{\nu}}{m_{Z}}\right)^{2}\right]^{3/2} \,\;  \theta(m_Z-2\,m_{\nu}).
\label{eq:Zlr}
\end{equation}
%Sizeable mixings reduce the coupling to the $Z$--boson, which couples only to LH fields, thus having an impact on all the neutrino phenomenology.
A further contribution should in principle be considered, that is the left - right mixing of the $\sn$, which could also participate in the Z decay. However, it becomes important only when the mass of the $\sn$ is lighter than $m_Z/2$.

\section{Sneutrino as Dark Matter candidate}
The models considered in this paper are by construction very similar to the MSSM. In the ordinary MSSM the LSP is the neutralino in extended regions of the parameter space, which anyway provides an interesting candidate of dark matter. However our interest here is to study another dark matter candidate, which naturally arises in this simple extensions of the MSSM, which is the lightest of the supersymmetric partners of the RH neutrinos, the $\sn$. It is an interesting candidate too, since it has electroweak scale interactions, and because of the mixing between the LH and RH components, does not suffer from the problems derived from a too strong coupling to the Z boson like the LH sneutrino of the MSSM.\\
The existence of the RH sneutrino as LSP does not exclude the possibility of having neutralino as LSP, though; the mass of the LSP strongly depends on the new parameters added to the theory. There will then be just some regions of the parameter space where the RH sneutrino is the LSP.
\subsection{Model I with inverse seesaw}
The $\sn$ has a direct coupling to the Higgs, as can be seen from the term $Y_\nu \nu l H_u$ of the superpotential Eq. \ref{eq:ModIsupp}. This coupling depends on the Yukawa coupling $Y_\nu$ and on the trilinear term $T_{Y_\nu}$ and provides tree level interactions with the SM particles at the electroweak scale, thus making the $\sn$ a possible WIMP candidate. The strongest constraint on  $\sn$ to be a viable CDM candidate, comes from the CMB measuraments, that is from its relic abundance (RA).\\
In order to perform the analysis of the sneutrino dark matter phenomenology,  we have created a suitable model file for this model with \texttt{SARAH} \cite{Staub:2008uz}. Then, we have used \texttt{SPheno} \cite{Porod:2003um} to compute the low energy spectrum through a precise mass calculation using two-loop RGEs and one-loop corrections to all masses.  The calculation of the relic density of the LSP is then done with \texttt{MicrOmegas} \cite{Belanger:2006is} version \texttt{2.4.5} based on the \texttt{CalcHep} output of \texttt{SARAH}. To perform the scans we  used a Mathematica package (\texttt{SSP}) \cite{Staub:2011dp}.\\

\begin{figure}[htb]
\centering
{\includegraphics[scale=0.55]{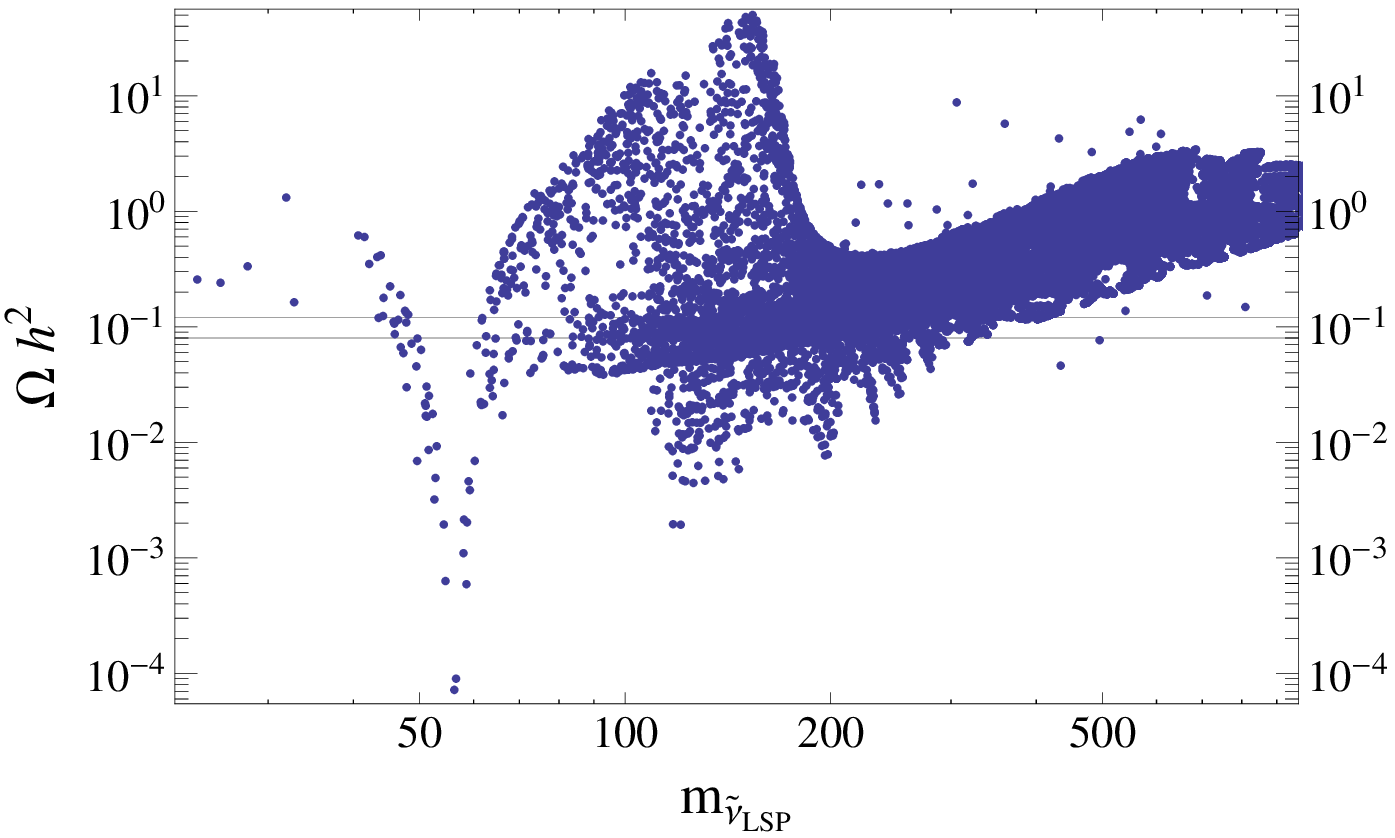}}
{\includegraphics[scale=0.55]{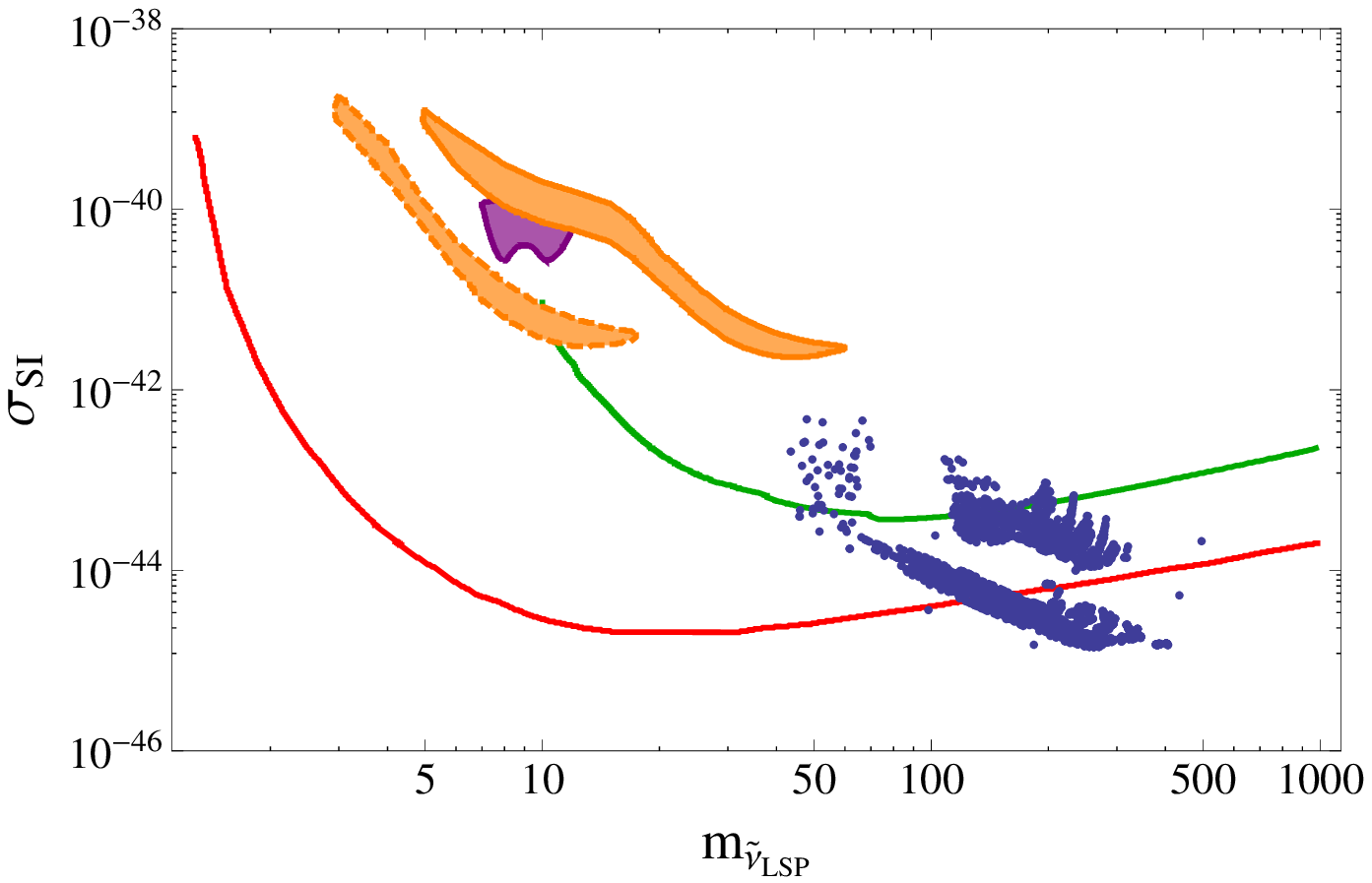}}  
\caption{Fig. a (left): General scan for Model I, showing the RA versus the $\sn$ mass. Fig. b (right): Direct detection cross section for model I, and  current limits from XENON100 \cite{Aprile:2012vw} (red line), CDMS \cite{Ahmed:2009zw} (green line), DAMA (with and without channeling, orange regions) \cite{Bernabei:2010mq}, and Cogent \cite{Aalseth:2010vx} (purple region).}
\label{fig:gen}
\end{figure}

The main annihilation channels are: 
\begin{itemize}
\item $\sn \sn \rightarrow W^+ W^-, Z^0 Z^0, f \bar{f}$ via s-channel Higgs ($h^0, H^0, A^0$) exchange;
\item  $\sn \sn \rightarrow W^+ W^-,  Z^0 Z^0$ via t- and u- channel $\tilde{l}$ exchange, and through a quartic coupling;
\item $\sn \sn \rightarrow H^0 H^0 $ through a quartic coupling which results to be very efficient;
\end{itemize}
In Fig.\ref{fig:gen}, where the $\sn$ is mainly LH, it behaves like a MSSM sneutrino. When its mass approaches half the mass of the Higgs, the "Higgs pole" is  clearly visible, which is due to the opening of the annihilation channel through the Higgs boson. This channel results to be highly efficient, indeed the RA goes down to very small values of the order of $10^{-5}$. Those points refer the sneutrino could not account for the totality of the CDM relic abundance, but it would still be a valid subdominant dark matter candidate. The analogue Z pole is instead suppressed, because of the dependence of the RA on $Y_\nu$ ($\propto Y_\nu^4$) and because of helicity. Another efficient channel it the quartic coupling to the Higgses, which appears when the mass of the $\sn$ is larger than the mass of the Higgs.  

\subsubsection{Direct Detection}

Direct detection of the sneutrinos would consist in detecting a signal coming from their elastic scattering with nuclei inside a detector. The direct detection rate of the DM particle depends on its  mass and on  the scattering cross section. The interaction, which occurs in a non relativistic limit, provided that the velocity of dark matter particles in the galactic halo is small, consists of basically two diagrams contributing at tree level, which are the t-channel exchange of a neutral Higgs or of the Z boson. This last diagram is however suppressed because of the mostly singlet nature of the  $\sn$.\\
Lepton number violation in the sneutrino mass matrix leads to a mass splitting between the real and the imaginary part of the lightest sneutrino, and the scattering via Z boson exchange occurs inelastically, through a transition from the real to the imaginary or viceversa. If the mass splitting is greater than some keV, scattering is strongly suppressed at direct detection experiments.\\
%Indeed, the maximum  kinetic energy that the sneutrino LSP can transfer to the detector depends on  the velocity it moves relative to the nucleus $v$ ($\simeq 10^-3$ in the galactic halo),the nucleus mass M and the angle $\theta$ of scattering:
%
%\begin{equation}
%E = \frac{A^2 v^2}{M} (1-cos(\theta))
%\end{equation} 
%
%where $A = \frac{m_{\sn} M}{m_{\sn} + M}$, which would give, in the case of a Xenon detector for istance, and $m_{\sn} = 100$ GeV, E = 25 keV (if cos($\theta$) = 0). For heavier sneutrinos with a mass of the order of TeV, the splitting cannot be larger of some hundred keV.\\ 
In Fig.\ref{fig:gen}b we compare the direct detection cross section for the RH sneutrino LSP with current direct detection experiments. The major bound nowadays comes from the XENON100 experiment \cite{Aprile:2012vw}, whose best sensitivity is around $10^{-44} \rm cm^2$ for a dark matter candidate of 50 GeV. We depict the direct detection cross section versus the LSP sneutrino mass (blue points). The points are those which survive the relic abundance constraint. In the same plot, the current limits from XENON100 \cite{Aprile:2012vw} (red line), CDMS \cite{Ahmed:2009zw} (green line), DAMA (with and without channeling, orange regions) \cite{Bernabei:2010mq}, and Cogent \cite{Aalseth:2010vx} (purple region) are shown. The sneutrinos show a SI cross section 
$\sigma_{SI} \lesssim 10^{-42} \rm cm^2$, and for masses $m_{\sn} \gtrsim 100$ GeV they are compatible with current limits by XENON100. However, XENON1T, whose sensibility should improve up to $10^{-46} \rm cm^2$, will test those cross sections.\\

\subsection{Model I with linear seesaw}
The phenomenology of sneutrino dark matter in the MSSM with linear seesaw is basically the same as in the previous case. Indeed, it is not possible to distinguish between the kind of seesaw mechanism through the dark matter phenomenology.\\
\subsection{Model II}
\begin{figure}[htb]
\centering
{\includegraphics[scale=0.55]{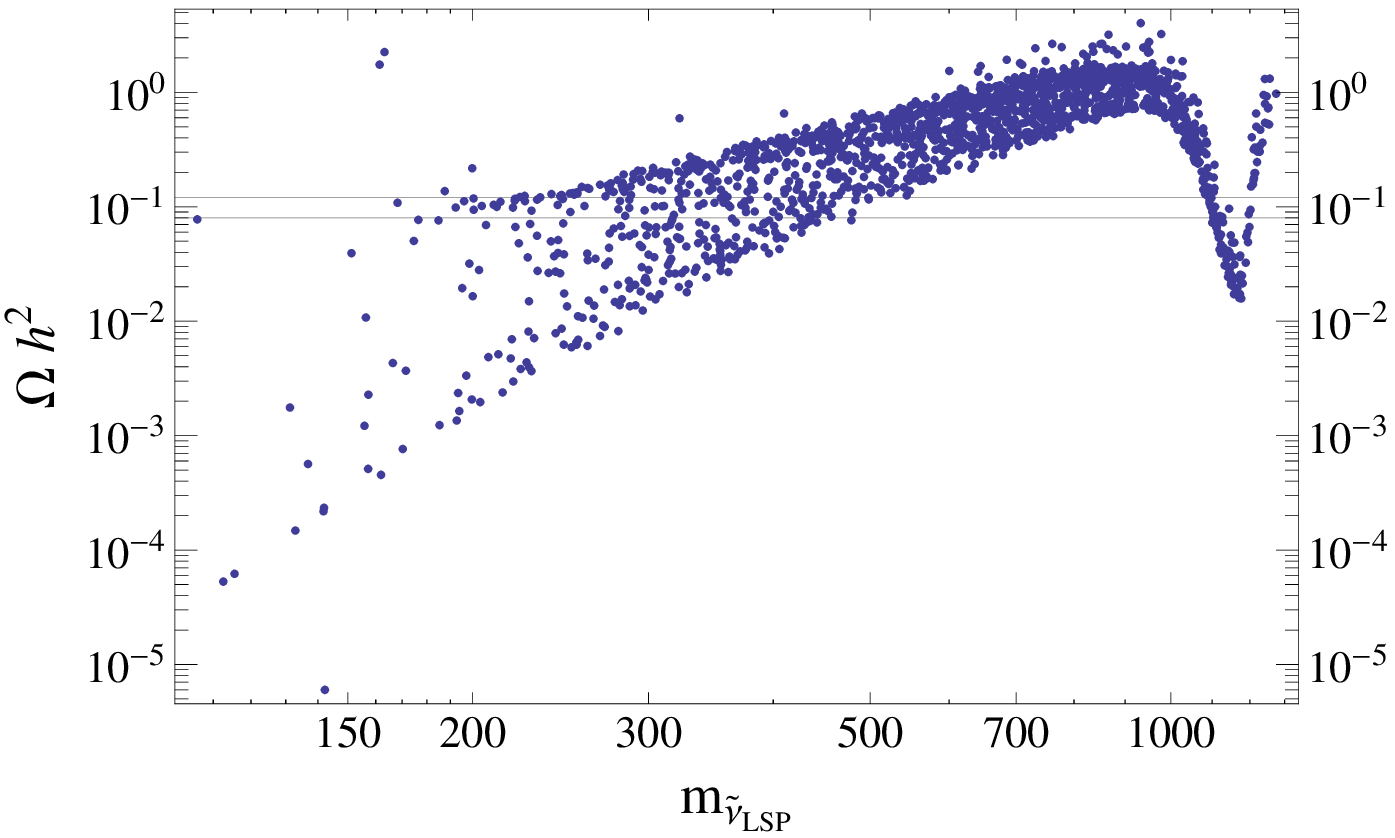}}
{\includegraphics[scale=0.55]{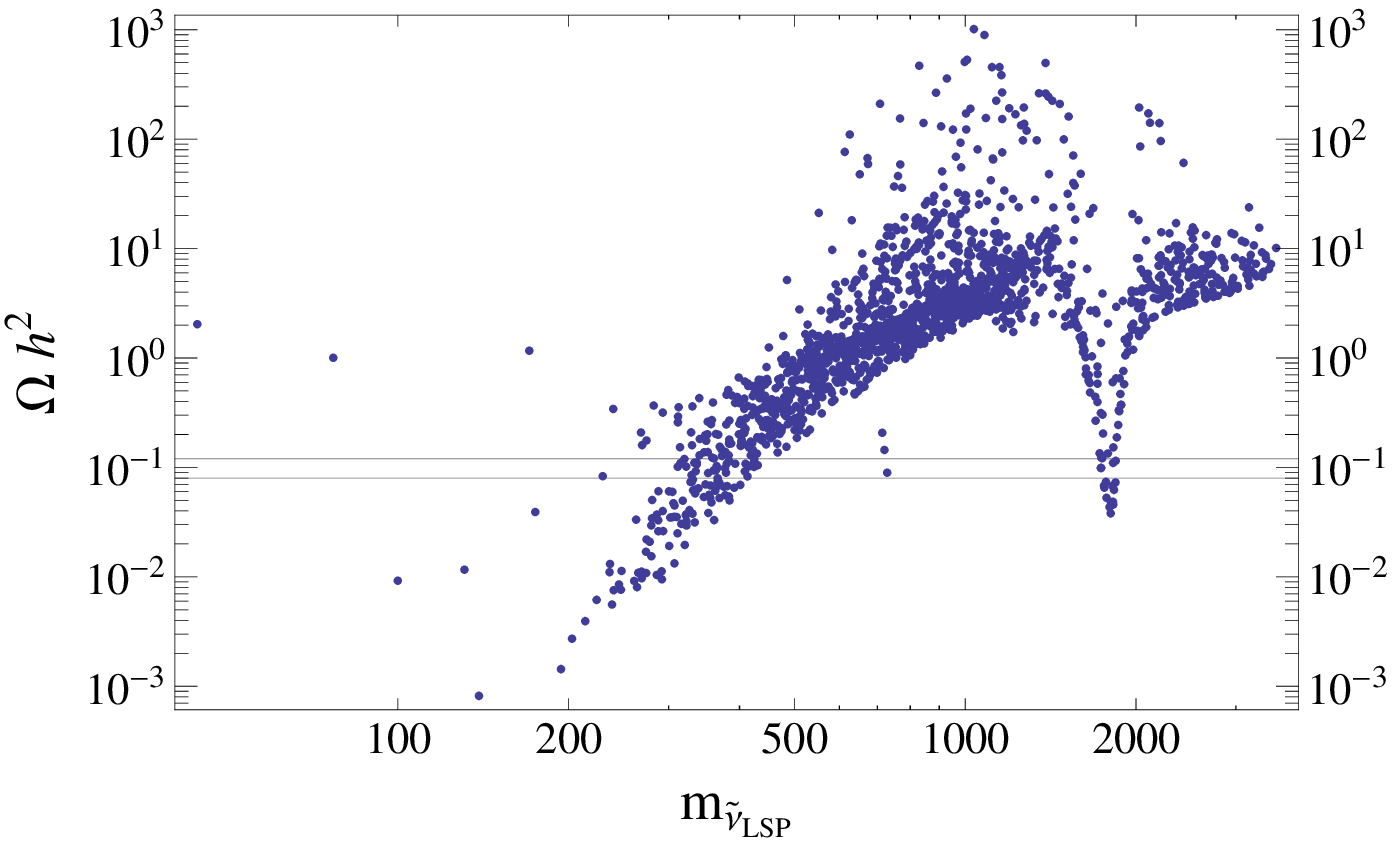}}
\caption{Scans for model II, made over the parameters $m_0 = [0,6000]$ GeV, $M_{1/2} = [3000,8000]$ GeV, tan$\beta_R$ = [1.0,1.3]. The other parameters are set to the values tan$\beta$=10, $A_0 = -4500$ GeV, $Y_S =$ diag(0.3); $v_R$ has been chosen different in the two plots,  $v_R = 5.5$ TeV (left) and  $v_R = 10$ TeV (right). }
\label{fig:BLRDMgenvR}
\end{figure}

In this subsection we discuss the DM phenomenology in the minimal supersymmetric $U(1)_R \times U(1)_{B-L}$ extension of the standard model. The presence of the extra gauge boson $Z'$  leads to important differences with respect to the previous models.\\
%%The $U(1)_{R}\times U(1)_{B-L}$ gauge symmetry of this model is spontaneously broken to the hypercharge group $U(1)_{Y}$ by the vevs $v_{\chi_{R}}$ and $v_{\bar\chi_{R}}$ of the scalar components of the $\hat\chi_R$ and $\hat{\bar{\chi}}_R$ superfields whereas the $SU(2)_{L}\otimes U(1)_{Y}\to U(1)_{Q}$ breakng is driven by the vevs $v_{d}$ and $v_{u}$ of the neutral scalar components of the $SU(2)_L$ Higgs doublets $H_d$ and $H_u$ up to gauge kinetic mixing effects.
%The tadpole equations for the different vevs can be solved analytically for either (i) ($\mu,B_\mu, \mu_R$, $B_{\mu_R}$) or (ii) ($\mu, B_\mu$, $m^2_{\chi_R}$,$m^2_{{\bar\chi}_R}$) or (iii) ($m^2_{H_d}$, $m^2_{H_u}$, $m^2_{\chi_R}$, $m^2_{{\bar\chi}_R}$) \cite{Hirsch:2012kv}. We address the minimal version option (i) as CmBLR (constrained mBLR), since it allows to define boundary conditions for all scalar soft masses at $m_{GUT}$, reducing the number of free parameters by four, although leading to some constraints on the parameter space, such as a lower bound on tan$\beta_R$ (tan$\beta_R >1$) \cite{Hirsch:2012kv}. The second option (ii) is instead more flexible, and we will make use of it in some of our scans, to which we will refer as $\chi_R$mBLR version(non-universal $\chi_R$ masses mBLR). We will not use the last option, which we only mentioned for the sake of completeness.\\
We use a version of the model where the tadpole equations for the vevs are solved analytically for ($\mu,B_\mu, \mu_R$, $B_{\mu_R}$) and we perform a scan over the parameters $m_0$, $M_{1/2}$, tan$\beta_R$. The results are shown in Figg. \ref{fig:BLRDMgenvR}. 

In both plots the main feature clearly visible is the $Z'$ pole. Indeed, the annihilation of the $\sn$ LSPs into SM particles via the $Z'$ becomes efficient when the mass of the $\sn$ is close to half the mass of the $Z'$. The mass of the $Z'$  can be calculated analytically \cite{Hirsch:2012kv} and mainly depends on the value of $v_R$. The ATLAS searches for a $Z'$ set a lower limit on its mass which is 1.8 TeV, and this translates into a lower limit on  $v_R \gtrsim 5$ TeV. In Fig. \ref{fig:BLRDMgenvR} we chose two different values of $v_R$: $v_R = 5.5$ TeV and  $v_R = 10$ TeV, thus leading to two values of $Z'$ and so to a shift in the position of the pole in the relic density plot. This shows that choosing a higher value of  $v_R$ we can get heavy $\sn$ DM with the correct RA.\\
The quartic coupling with two Higgses ( $h^0$, $h^0_{BLR}$ and $A^0$, depending on if they are kinematically allowed, according to the $\sn$ mass) is one of the most efficient annihilation channels, as before. For lower masses the annihilation via the MSSM Higgs is the most efficient, as can be noticed by the lowering of the relic density going to lower masses, expecially in the first plot, where on the left end side we are approaching the Higgs pole.

\subsubsection{Direct Detection}
\begin{figure}[hbt]
\centering
\includegraphics[scale=0.55]{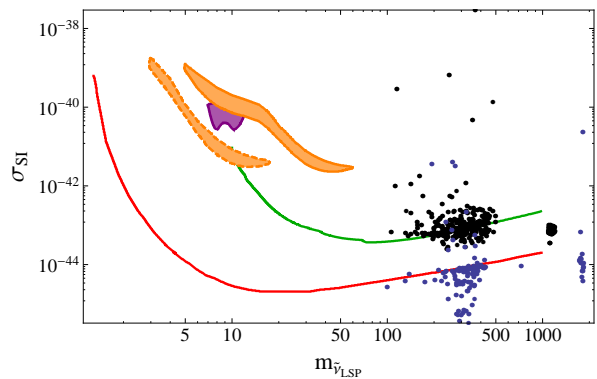}  
\caption{Direct Detection cross section limits for model II. The two clouds of points refer to  $v_R = 5.5$ TeV (black) and $v_R = 10$ TeV (blue). Experimental limits as in Fig. \ref{fig:gen}b. }
\label{fig:BLRDD}
\end{figure}

As to the direct detection cross section analysis, a  bound on the $v_R $, and then on the mass of the $Z'$ can be set by the DD experiments, expecially by Xenon100. This bound turns to be stronger than the one set by the collider experiments like LHC, as it is shown in Fig.\ref{fig:BLRDD}. The two clouds of points there refer to  $v_R = 5.5$ TeV (black) and $v_R = 10$ TeV (blue).

\section*{Acknowledgments}
V.D.R. warmly thanks Martin Hirsch, with whom this work was done, for many useful comments. The work of V.D.R. is supported by the EU~Network grant UNILHC PITN-GA-2009-237920. We also acknowledge support from the Spanish MICINN grants FPA2011-22975 and MULTIDARK CSD2009-00064 (Con-solider-Ingenio 2010 Programme) and by the Generalitat Valenciana grant Prometeo/2009/091.
%\end{acknowledgments}

\section*{References}%%%%%%%%%%%%%%%%%%%%%%%%%%%%%%%%%%%%%%%%%%%%%%%%%%%%%%%%%%%%%%%%%%%%%%%
%%%%%%%%%%%%%%%%%%%%%%%%%%%%%%%%%%%%%%%%%%%%%%%%%%%%%%%%%%%%%%%%%%

\end{document}